\documentclass[aps,pra,twocolumn,superscriptaddress,nofootinbib]{revtex4-1}

\usepackage{amsmath,amssymb,amsfonts}
\usepackage{graphicx}
\usepackage{dcolumn}
\usepackage{bm}
\usepackage{hyperref}
\usepackage[sort&compress]{natbib}
\usepackage{color}
\usepackage{verbatim}
\usepackage{braket}
\usepackage{todonotes}
\usepackage{mathrsfs}
\usepackage{extarrows}

\newcommand{\mbf}{\mathbf}
\newcommand*{\dif}{{\,\rm d}}
\newcommand*{\p}{\mathop{}\!\mathrm \partial}

\usepackage{mathtools}

\begin{document}

\title{Universal scaling of unequal-time correlation functions in ultracold Bose gases\\ far from equilibrium}

\author{Andreas Schachner}
   \email{schachner@thphys.uni-heidelberg.de}
\author{Asier Pi\~neiro Orioli}
   \email{pineiroorioli@thphys.uni-heidelberg.de}
\affiliation{Institut f\"ur Theoretische Physik, Universit\"at Heidelberg, Philosophenweg 16, 69120 Heidelberg, Germany}
\author{J\"urgen Berges}
   \email{berges@thphys.uni-heidelberg.de}
\affiliation{Institut f\"ur Theoretische Physik, Universit\"at Heidelberg, Philosophenweg 16, 69120 Heidelberg, Germany}

\begin{abstract}
We explore the far-from-equilibrium dynamics of Bose gases in a universal regime associated to nonthermal fixed points. While previous investigations concentrated on scaling functions and exponents describing equal-time correlations, we compute the additional scaling functions and dynamic exponent $z$ characterizing the frequency dependence or dispersion from unequal-time correlations. This allows us to compare the characteristic condensation and correlation times from a finite-size scaling analysis depending on the system's volume.   
\end{abstract}

\maketitle

\section{Introduction}
\label{sec:intro}

Nonequilibrium scaling phenomena are ubiquitous in nature. A particularly well-understood example concerns dynamic scaling behavior near second-order thermal phase transitions, where a wide variety of physical systems can be grouped into universality classes associated to thermal renormalization group fixed points. Each universality class is characterized by a set of values of critical exponents and scaling functions describing the long-distance properties of systems~\cite{hohenberg1977theory}. 

While quenches to second-order thermal phase transitions can be still characterized by the universal critical behaviour, quenches across transitions are typically well described in terms of the phenomenon of coarsening: on the low-temperature side the systems form domains which grow with time such that correlation functions can be expressed in terms of scaling functions and power laws. Their forms and values depend on the condensate structure and topological obstructions. Therefore, the study of topological defects provides a case-by-case framework for discussing coarsening in these different systems~\cite{Bray:1994kin}.

More recently, new universality classes associated to nonthermal fixed points have been discovered in the context of thermalization dynamics in the early universe after inflation~\cite{Berges:2008wm,Berges:2008sr,Micha:2004bv}, heavy-ion collisions described by quantum chromodynamics~\cite{Berges:2013eia,Berges:2013lsa,Kurkela:2015qoa} and setups with ultracold quantum gases~\cite{Scheppach:2009wu,nowak2012nonthermal,nowak2011superfluid,Orioli:2015dxa,karl2016strongly}. The scaling behavior of these initially over-occupied systems is described in terms of universal exponents and scaling functions. The latter are self-similar attractor solutions to which the system evolves without fine-tuning of any relevant operator.   
Nonthermal fixed points can characterize remarkably large universality classes, encompassing relativistic and non-relativistic quantum and classical theories even with different symmetries and field content~\cite{Berges:2014bba,Orioli:2015dxa}.

So far, the most detailed understanding of nonthermal fixed points has been obtained for the dynamics of scalar fields with $N$ components. While dynamic properties at shorter distances can be related to the phenomenon of weak wave turbulence~\cite{Micha:2004bv}, the long-distance scaling behavior is reminiscent of ordering dynamics with the phenomenon of condensate formation~\cite{Berges:2012us,Orioli:2015dxa,Davis:2016hwt,SvistunovPhysRevA.66.013603,SachdevPhysRevA.54.5037}. In contrast to expectations from coarsening, the infrared scaling behavior is described in terms of universal exponents and scaling functions that are remarkably insensitive to the condensate structure and topological considerations~\cite{Berges:2010ez,Orioli:2015dxa}. As a consequence, important aspects of these phenomena can be described using large-$N$ expansions beyond leading order~\cite{Orioli:2015dxa,Berges:2016nru,Berges:2001fi}, which do not capture topological defects~\cite{Rajantie:2006gy,Rajantie:2010tb,Berges:2010nk}. The universality accross the wide set of condensate structures for different values of $N$ has been scrutinized in Ref.~\cite{Moore:2015adu} using classical-statistical simulations for the relativistic $N$-component field theory. The latter has been demonstrated in Ref.~\cite{Orioli:2015dxa} to be also in the same universality class as its non-relativistic counterpart. 

Here we extend previous work on nonthermal fixed points by providing first results on the universal scaling of unequal-time correlation functions. The latter give direct access to the important ``dynamic" scaling exponent $z$, which describes the characteristic frequency dependence or dispersion in the scaling regime. While close to thermal equilibrium, the dynamic exponent $z$ may also be inferred from equal-time correlations using scaling relations~\cite{hohenberg1977theory}, this is less clear far from equilibrium. For instance, in scaling regimes for energy transport towards short distance scales $z$ represents an independent exponent~\cite{Micha:2004bv}. To establish the universality classes of nonthermal fixed points, it is therefore crucial to determine the role and value of $z$. 

More precisely, we extract all scaling properties of two-times correlation functions for a Bose gas described by non-relativistic complex scalar fields in three spatial dimensions. The Bose gas corresponds to an $O(N)$ symmetric system for $N=2$ real scalar field components. The universal exponents and scaling functions are obtained from a finite-size scaling analysis depending on the system's volume. In particular, this allows us to compare the effective condensation and correlation times for finite systems by establishing their power-law scaling with volume.    

For the numerical simulations we exploit the fact that the quantum- and classical-statistical systems belong to the same universality class because of the large characteristic occupancies involved~\cite{Orioli:2015dxa,Berges:2016nru}. The comparisons to analytic estimates are based on extrapolations of large-$N$ results at next-to-leading order~\cite{Orioli:2015dxa} to the case $N=2$ considered. 

In section \ref{sec:scalingansatz} we describe the finite-size scaling ansatz for two-times correlation functions. Section~\ref{ssec:evol} presents a class of initial conditions characterizing over-occupied systems and their time evolution. Universal exponents and scaling functions are determined in section~\ref{sec:selfsim_regime}, where we also comment on the value of the anomalous dimension. The conclusions are given in section~\ref{sec:conclusions}. 
We explain our fit routines with error estimates in appendix~\ref{sec:FitProcedure}.

\section{Scaling of nonequilibrium correlation functions}\label{sec:scalingansatz}

We consider non-relativistic Bose gases out of equilibrium, whose quantum many-body dynamics may be described in terms of a complex bosonic Heisenberg field operator $\psi(t, \mbf x)$. The nonequilibrium evolution is encoded in correlation functions of fields at different space-time points. We investigate spatially homogeneous systems such that, for instance, the two-point correlation function of the anti-commutator 
\begin{equation}
	F(t,t^\prime, \mbf x - \mbf x^\prime) \equiv \frac{1}{2}\langle \psi(t,\mbf x)\psi^\dagger(t^\prime,\mbf x^\prime) + \psi^\dagger(t^\prime,\mbf x^\prime) \psi(t,\mbf x) \rangle 
\label{eq:anticom}
\end{equation}
depends only on the spatial difference $\mbf x-\mbf x'$, while the nonequilibrium evolution entails a breaking of time-translation invariance and a dependence on both $t$ and $t^\prime$ separately. 

For quantum systems, the brackets $\langle \ldots \rangle$ in (\ref{eq:anticom}) denote the quantum-statistical expectation value involving the trace over the density operator specifiying the initial state. We will concentrate on a range of far-from-equilibrium initial conditions involving large occupancies of typical modes, such that the quantum-statistical evolution can be accurately mapped onto a classical-statistical field theory problem to be simulated on a computer~\cite{berges2007quantum}.\footnote{The approximate mapping is usually based on a sufficiently large occupancy of typical modes for equal-time correlation functions. While the validity of this argument is less clear for the computation of general unequal-time correlation functions, it should be valid for power-law behavior in scaling regimes as considered in this work.} In classical-statistical simulations, one samples over initial conditions and evolves each realization according to the classical field equation of motion. In this case, the brackets $\langle \ldots \rangle$ denote the ensemble average over classical trajectories.

For both quantum and classical representations, at equal space-time points ($t = t^{\prime}$, ${\mbf x} = {\mbf x^\prime}$) the quantity (\ref{eq:anticom}) corresponds to the density $n$, i.e.~the conserved total particle number $N_{\rm total}$ divided by the system's volume $V$:
\begin{equation}
	n \equiv \frac{N_{\rm total}}{V} = F(t,t, \mbf x - \mbf x) \, .
\label{eq:numberconservation} 
\end{equation}
For our purposes, it is instructive to consider the absolute value of the spatial integral of (\ref{eq:anticom}) over the volume~$V=L^d$ in a box of length $L$ in $d$ spatial dimensions: 
\begin{equation}
	F(\tau, \Delta t, V) \equiv \left| \int_V  F(t,t^\prime, \mbf x) \, d^d x\, \right| .
\label{eq:n_def_stat_nonrel}
\end{equation}
Here we introduced the central-time coordinate $\tau$ and the relative-time coordinate $\Delta t$ as
\begin{equation}
	\tau\equiv\frac{t+t'}{2} \ , \quad \Delta t\equiv t-t' \, .
\label{eq:tau_dt}
\end{equation}
Taking the absolute value in (\ref{eq:n_def_stat_nonrel}) amounts to disregarding a rotating global phase $\sim e^{i\mu \Delta t}$, which could also be absorbed in a redefinition of the fields by $\psi \rightarrow e^{-i\mu \Delta t} \psi$. Furthermore, since the correlation function (\ref{eq:n_def_stat_nonrel}) is symmetric under exchange of $t$ and $t^\prime$, $F(\tau, \Delta t, V) = F(\tau, -\Delta t, V)$, we restrict our presentation to $\Delta t > 0$. Though we keep the dimension $d$ general in our notation, all our numerical results presented in subsequent sections will concern $d = 3$.

Our aim is to investigate scaling solutions of nonequilibrium correlation functions near nonthermal fixed points. The scaling behavior of the correlation function (\ref{eq:n_def_stat_nonrel}) may be expressed in terms of real scaling exponents $\alpha$, $\beta$ and $z$ as
\begin{equation}
F(\tau,\Delta t,V) = s^{\alpha/\beta} F(s^{-1/\beta} \tau, s^{-z}\Delta t, s^{-d} V)  
\label{eq:Fscaling}
\end{equation}
under rescaling with the real scaling parameter $s>0$. The ``occupation number" exponent $\alpha$ and the ``central-time" exponent $\beta$ have been discussed in detail for $\Delta t \equiv 0$ in Ref.~\cite{Orioli:2015dxa} for the far-from-equilibrium case we are interested in. To determine the ``dynamic" scaling exponent $z$, which is associated to changes in relative times $\Delta t$, and the respective unequal-time scaling functions is the main focus of our investigation.\footnote{The nonthermal scaling exponents $\alpha$ and $\beta$ are not associated to a specific heat or order parameter exponent but defined by (\ref{eq:Fscaling}).} 

Far from equilibrium, i.e.,~well beyond the linear response regime, both the central-time exponent $\beta$ and relative-time exponent $z$ can generally be linearly independent as, for instance, realized in perturbative scaling regimes for energy transport towards short distance scales in related models~\cite{Micha:2004bv}. In this work, we consider the nonperturbative scaling regime associated to particle transport towards long-distance scales~\cite{Orioli:2015dxa} and determine the role and value of $z$.

The importance of $z$ stems from the fact that the dynamic scaling exponent for relative times is directly related to the characteristic frequency dependence or dispersion of the model. The dependence on the frequency $\omega$ is obtained from Fourier transforming (\ref{eq:Fscaling}) with respect to relative times, which gives
\begin{eqnarray}
	F(\tau, \omega, V) &=& 2 \int_0^\infty e^{i\omega \Delta t}\, F(\tau, \Delta t, V)\, d(\Delta t)
	\nonumber\\
	&=& s^{z+\alpha/\beta} F(s^{-1/\beta} \tau, s^{z}\omega, s^{-d} V) \, .  
\label{eq:Fomega}
\end{eqnarray}

Because the system is considered to have a finite size $L^d$, when the characteristic correlation length is $\approx L$, the system can already become effectively ordered. Only for shorter times the universal scaling behavior with a full dependence on $\tau$, $\Delta t$ and $V$ as in (\ref{eq:Fscaling}) is expected to hold. Below we determine the corresponding time scale for condensation from equal-time correlation functions and compare this to the characteristic correlation time obtained from unequal-time functions.

\section{Initial conditions and nonequilibrium evolution}
\label{ssec:evol}

We envisage an interacting Bose gas in three spatial dimensions with s-wave scattering length $a$ and average density $n$. We focus on the dilute regime, such that the dimensionless parameter $\zeta = \sqrt{n a^3} \ll 1$. We think of preparing the system in an extreme nonequilibrium situation, where the typical occupation numbers are very much larger than in thermal equilibrium. As a consequence, the dynamics will be non-perturbative despite being in a dilute regime. 

To describe this extreme condition, we exploit the fact that
the density and scattering length can also be used to define a characteristic ``coherence length", whose inverse is described by the momentum scale $Q = \sqrt{16 \pi a n}$.\footnote{We always employ natural units where the reduced Planck constant and Boltzmann's constant are set to unity: $\hbar=k_B=1$.} To observe the dynamics near nonthermal fixed points for the interacting Bose gas, an unusually large occupancy of modes at the inverse coherence length scale $Q$ has to be prepared~\cite{Orioli:2015dxa}. Decomposing 
\begin{equation}
n = |\psi_0|^2 + V^{-1} \sum_{\mathbf p} f_0({\mathbf p})
\end{equation}
into a condensate fraction $|\psi_0|^2$ and non-condensate fraction with momentum distribution function $f_0({\bf p})$, we initially set $|\psi_0|^2 = 0$ and 
\begin{equation}\label{eq18}
f_0(Q)\, \sim \, \dfrac{1}{\zeta} \, \gg \, 1
\end{equation}
to describe highly occupied modes with typical momentum $Q$. In this case the large occupation number $\sim 1/\zeta$ compensates for the smallness of the diluteness parameter $\zeta$: the system becomes strongly correlated and independent of the value of $\zeta$~\cite{Orioli:2015dxa}. 

In particular, Bogoliubov or mean-field-type approximations are not applicable in this regime and we employ classical-statistical lattice simulations. More specifically, we compute correlation functions from an ensemble average of inhomogeneous solutions of a complex Bose field $\psi(t, \mbf x)$, whose dynamics is described by the Gross-Pitaevskii equation~\cite{gross1961structure}
\begin{equation}
i \p_t \psi(t, \mbf x) = \left (-\dfrac{\nabla^2}{2m}+g\, |\psi(t, \mbf x)|^{2}\right ) \psi(t,\mbf x)
\label{eq:gpe}
\end{equation}
with mass $m$ and interaction parameter $g=4\pi a /m$. With $Q= 2 \sqrt{n m g}$ and $\zeta = mgQ/(16 \pi^{3/2})$, we sample the fields at initial time such that
\begin{equation}
f_0({\mathbf p}) := \int_V d^3x e^{-i{\mbf p}{\mbf x}} \frac{1}{2} \left\langle \psi(0,{\mbf x}) \psi^\dagger(0,{\mbf 0}) + \psi^\dagger(0,{\mbf 0}) \psi(0,{\mbf x}) \right\rangle 
\end{equation}
is given by $f_0(\mathbf{p})=25/(mgQ)$ for momenta $|\mathbf{p}| < Q$ and zero otherwise.

To reflect the classical-statistical nature of the dynamics in the highly occupied regime, we measure time in units of $2m/Q^2$ and volumes in units of $Q^3$. As a consequence, the combination $F(\tau, \Delta t, V) 2mgQ$ for (\ref{eq:n_def_stat_nonrel}) does not depend on the values of $m$, $g$ and $Q$. Though we will write $t$, $V$ and $F$, we always imply the rescalings $t\rightarrow t Q^{2}/2m$, $\mathbf{p}\rightarrow \mathbf{p}/Q$, $V\rightarrow VQ^{3}$ and $F(\tau,\Delta t ,V)\rightarrow F(\tau,\Delta t,V)\, 2mgQ$ in the following.

\begin{figure}[t!]
\centering
\includegraphics[width=.5\textwidth]{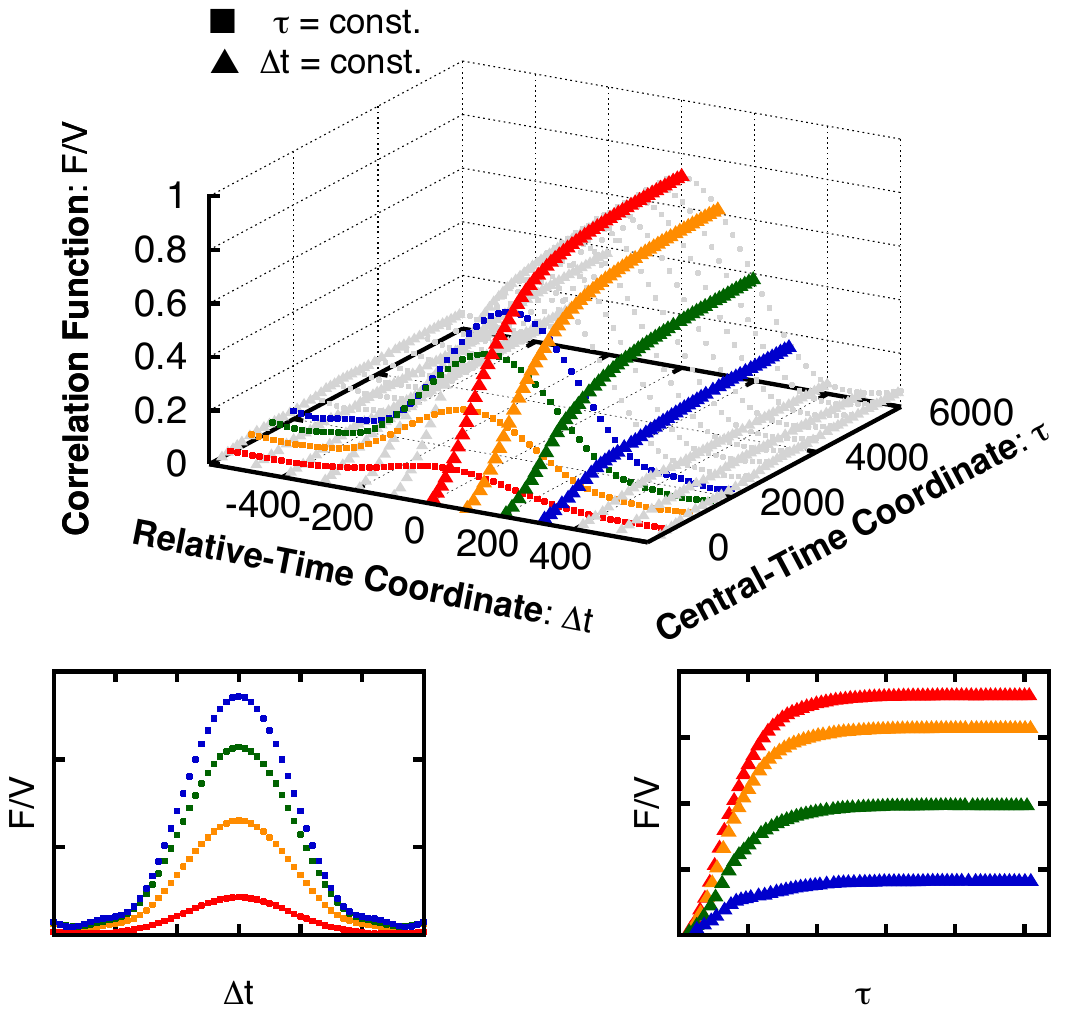}
\caption{The upper graph shows the two-times correlation function (\ref{eq:n_def_stat_nonrel}) as a function of the central-time coordinate $\tau$ and the relative-time coordinate $\Delta t$ for a volume $V$ with $128^{3}$ lattice points. The colored lines correspond to slices of constant $\tau$ (squares) and constant $\Delta t$ (triangles), which are separately displayed in the lower graphs.}\label{fig:overview_128}
\end{figure}

To give an overview, Fig.~\ref{fig:overview_128} shows the evolution of the correlation function $F(\tau,\Delta t, V)$ as a function of the central-time coordinate $\tau$ and the relative time $\Delta t$ for a volume $V=128^3$.\footnote{For all numerical estimates we employ an ultraviolet cutoff at $\sqrt{12}\,Q$.} For better visualization, the lower graphs of Fig.~\ref{fig:overview_128} give slices of constant $\tau$ ($\Delta t$) as a function of $\Delta t$ ($\tau$) in the left (right) plot. 

The decay of $F$ as a function of $\Delta t$ establishes a characteristic correlation time $\Delta t_*(V)$, whose scaling with volume is investigated in detail in section~\ref{sec:selfsim_regime}. Likewise, the growth of $F$ as a function of $\tau$ is seen to terminate around a time $\tau_*(V)$, which is discussed in the next section. In Ref.~\cite{Orioli:2015dxa}, $\tau_*(V)$ has been associated to the characteristic time scale for condensate formation.

\section{Extracting universal exponents and scaling functions}\label{sec:selfsim_regime}

In a scaling regime described by (\ref{eq:Fscaling}), we may choose \mbox{$s = V^{1/d}$} eliminating the scaling parameter to obtain
\begin{equation}
 F(\tau ,\Delta t,V) \,= \, V^{\alpha/(\beta d)}\, F_V(V^{-1/(\beta d)} \tau, V^{-z/d}\Delta t)  \, ,
\label{eq:FscalingV}
\end{equation} 
where the scaling function $F_V$ is defined in terms of
$F_V(V^{-1/(\beta d)} \tau, V^{-z/d} \Delta t)\equiv  F(V^{-1/(\beta d)} \tau, V^{-z/d} \Delta t,1)$. This form makes it explicit that in the scaling regime $F_V$ depends only on two arguments instead of separately on $\tau$, $\Delta t$ and $V$. 
Similarly, it is instructive to consider the choices $s=\tau^{\beta}$ in (\ref{eq:Fscaling}) leading to 
\begin{equation}
 F (\tau, \Delta t,V ) \, =\, \tau^{\alpha}\, F_\tau (\tau^{-\beta z} \Delta t, \tau^{-\beta d} V ) \, ,
\label{eq:Fscalingtau}
\end{equation}
or $s=\Delta t^{1/z}$ in (\ref{eq:Fscaling}) to get the scaling form
\begin{equation}
 F (\tau, \Delta t,V ) \, =\,  \Delta t^{\alpha/\beta z}\, F_{\Delta t}(\Delta t^{-1/\beta z}\, \tau,\Delta t^{-d/z} V ) \, .
\label{eq:Fscalingdt}
\end{equation}

One may use any of the scaling forms (\ref{eq:FscalingV})--(\ref{eq:Fscalingdt}) to efficiently extract the universal scaling exponents $\alpha$, $\beta$ and $z$ from the numerical data. The different shapes of the scaling functions $F_V$, $F_\tau$ and $F_{\Delta t}$ are also universal after fixing their overall amplitudes and of their arguments.

\begin{figure}[t!]
	\centering
	\includegraphics[width=.49\textwidth]{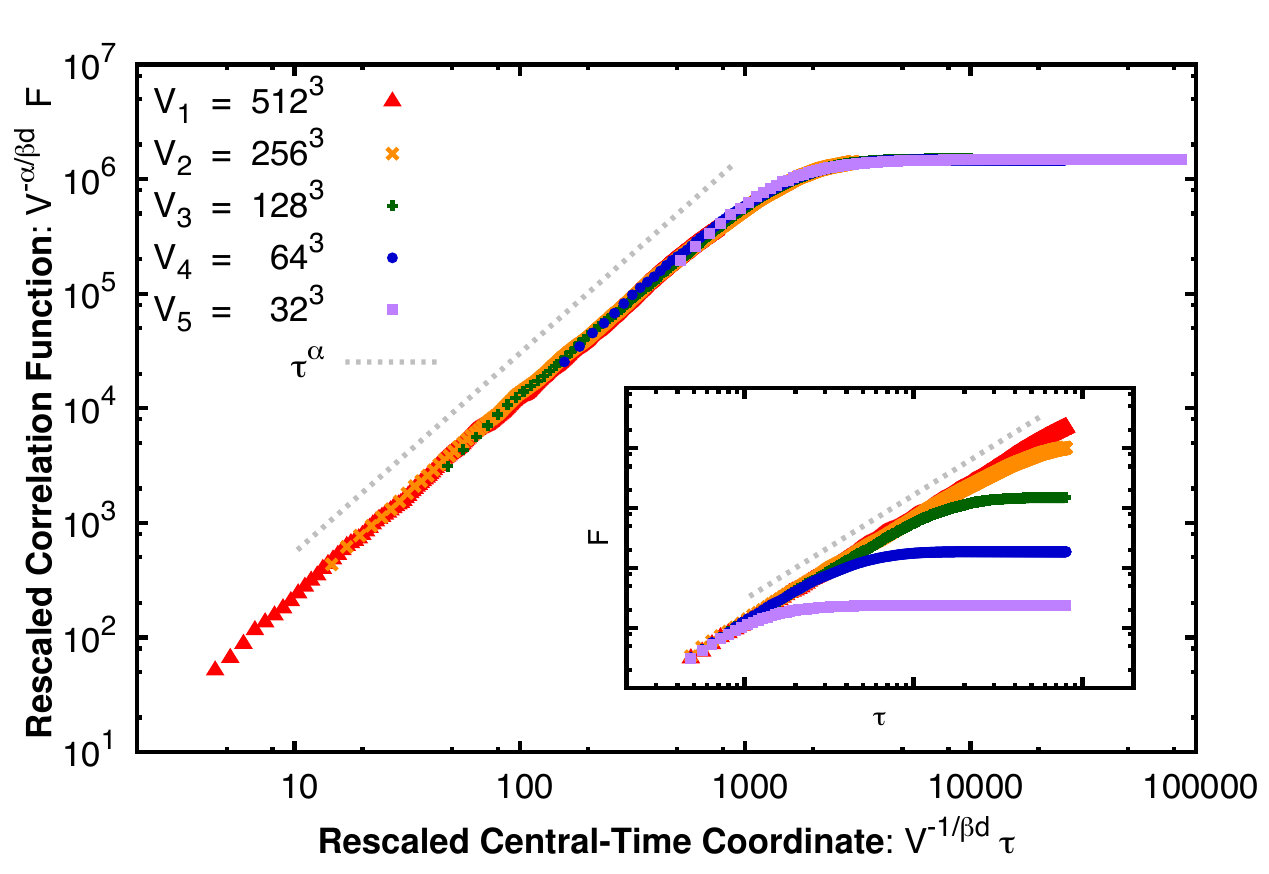}
	\caption{Rescaled correlation function $F_V=V^{-\alpha/(\beta  d)}F$ as a function of the rescaled central time $V^{-1/(\beta d)}\tau$ for $\Delta t=0$ and a range of volumes $V$ in $d=3$ spatial dimensions. For the rescalings we employ $\alpha/(\beta d)=1$ and $1/(\beta d)=0.57$. The inset shows the function $F$ without rescaling for comparison. The dashed line represents the power law behavior $\sim\tau^{\alpha}$ with $\alpha=1.74$. The time where the power-law behavior stops and the curve flattens indicates the characteristic condensation time $\tau_*(V)$.
		}
	\label{fig:selfsim_fit1}
\end{figure}

Because the system has a finite size $L^d$, it can already become effectively ordered at a finite time, which has been studied from equal-time correlations in Ref.~\cite{Orioli:2015dxa}. Using the scaling form (\ref{eq:FscalingV}), we denote the condensation time
\begin{equation}
\tau_* \, \sim \, V^{1/(\beta d)}
\label{eq:scalingtaustar}
\end{equation}
with $\tau_* = \tau_*(V,\Delta t = 0)$
as the time where $F_V$ at equal-times becomes approximately independent of $\tau$ for given volume $V$, i.e.,~$F_V(V^{-1/(\beta d)}\tau,0) \simeq \mbox{const}$ for $\tau \gtrsim \tau_*$. That $F_V$ changes its behavior qualitatively from a power-law \mbox{$\sim \tau^\alpha$} to become an approximate constant around the time $\tau_*$ is indeed seen in numerical solutions as demonstrated in Fig.~\ref{fig:selfsim_fit1}. The figure is discussed in more detail below when we extract the values of the scaling exponents. The interpretation of $\tau_*$ as the time for the formation of a Bose condensate is explained in Ref.~\cite{Orioli:2015dxa}.

\begin{figure}[t!]
	\centering
	\includegraphics[width=.49\textwidth]{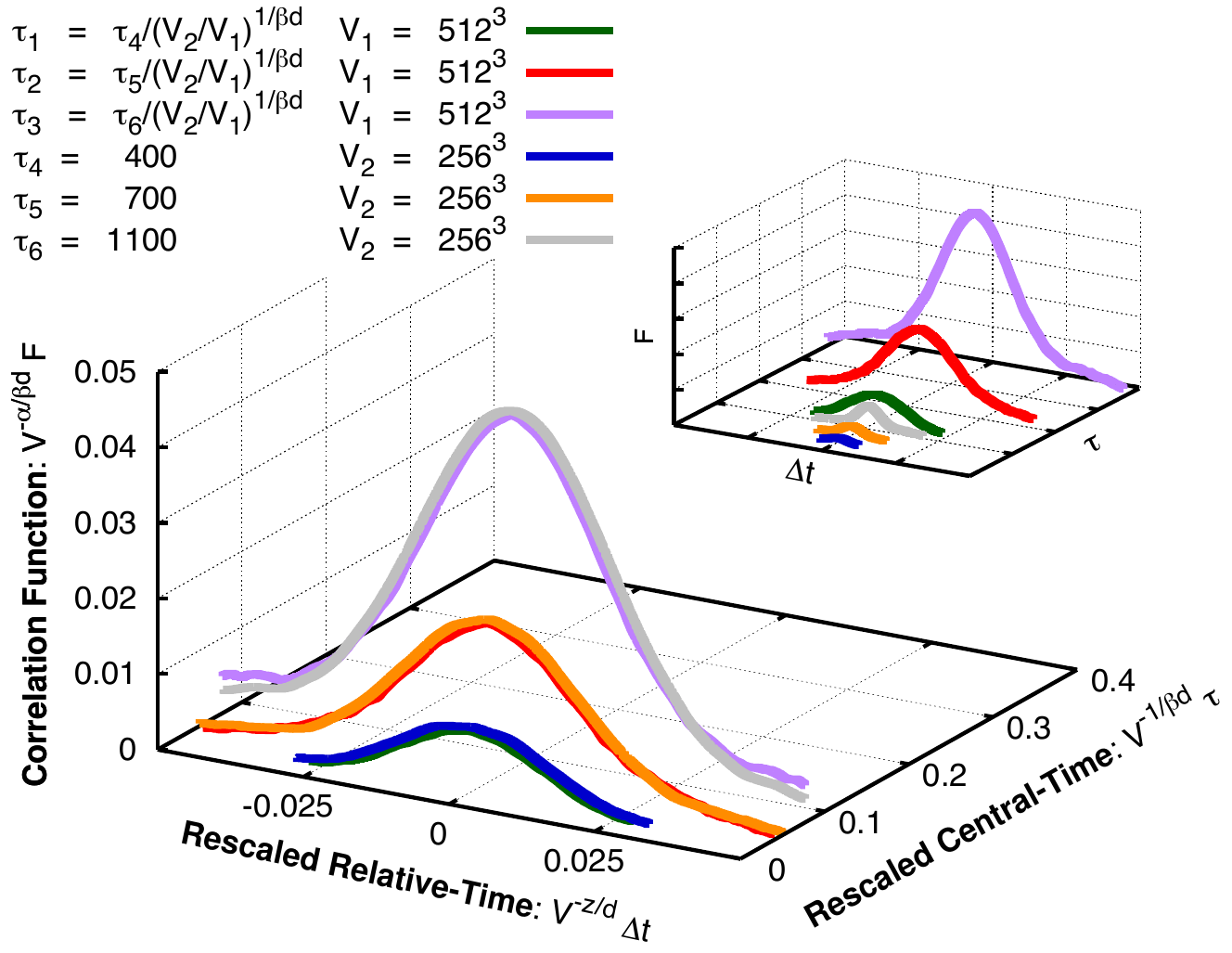}
	\caption{Rescaled correlation function $F_V=V^{-\alpha/(\beta d)}F$ as a function of the rescaled relative-time coordinate $V^{-z/d}\Delta t$ for fixed values of the rescaled central-time coordinate $V^{-1/(\beta d)}\tau$. We employ $\alpha/(\beta d)=1.0$, $1/(\beta d)=0.57$, and $z/d=0.61$ for two different volumes with $\{256^3,512^3\}$ lattice points. The inset shows the original function $F$ without rescaling. The ``width" of $F_V$ as it decays with growing $|V^{-z/d}\Delta t|$ gives rise to the characteristic correlation time $\Delta t_*(V)$.
		}
	\label{fig:selfsim_fit2}
\end{figure}

In addition, we define the correlation time $\Delta t_*$
from the decay of $F_V$ as a function of relative time, which is exemplified in Fig.~\ref{fig:selfsim_fit2}. More precisely, we determine the decay-time from the ``width" given by the difference between the inflection points of the curve \mbox{$F_V(V^{-1/(\beta d)}\tau=\mbox{const},V^{-z/d}\Delta t)$} as a function of $V^{-z/d}\Delta t$. This difference is found to grow monotonically with $\tau$ until it reaches a maximum at a time $\tau_\Delta(V)$, i.e.,~the width of the scaling function $F_V$ becomes independent of the central time for $\tau \gtrsim \tau_\Delta$. 

Though $\tau_\Delta$ and $\tau_*$ turn out to scale in the same way with volume as (\ref{eq:scalingtaustar}), they can be numerically different and we find $\tau_\Delta < \tau_*$. In particular, in this regime (\ref{eq:FscalingV}) implies
\begin{equation}
\Delta t_* \, \sim \, V^{z/d} 
\label{eq:scalingdeltatstar}
\end{equation}
with $\Delta t_* = \Delta t_*(V,\tau = \tau_\Delta(V))$.

Since $V = L^d$, the condensation time (\ref{eq:scalingtaustar}) and the correlation time (\ref{eq:scalingdeltatstar}) are related to respective lengths, which scale as  
\begin{equation}
L \, \sim \, \tau_*^\beta \, \sim \, \Delta t_*^{1/z} \, . 
\end{equation}
A special case occurs if $\beta = 1/z$ for which the scalings with central and relative times are the same. We analyze this possibility below.

In the following we extract the values of the universal exponents and determine the universal shape of the scaling functions. Starting from the initial conditions of section~\ref{ssec:evol}, we follow numerically the relatively short evolution until the system is attracted to the nonthermal fixed point characterized by scaling. 
We analyze the scaling behavior for times $\tau < \tau_\Delta(V)$ and $\Delta t < \Delta t_*(V,\tau_\Delta)$ for different volumes $V$. The evolution in this regime is verified to exhibit the scaling behavior (\ref{eq:Fscaling}) with suitably chosen exponents. 

We start by considering $\Delta t=0$ and plot the rescaled correlation function $F_V(V^{-1/(\beta d)} \tau, 0)$ as defined in (\ref{eq:FscalingV}). In Fig.~\ref{fig:selfsim_fit1} we show results for a set of volumes with $\{32^3,64^3,128^3,256^3,512^3\}$ lattice points, respectively. For comparison, the inset shows the correlation function $F(\tau,\Delta t=0,V)$ for the corresponding values of $\tau$ without rescaling. With the appropriate choice of values for the combinations of exponents $\alpha/(\beta d)$ and $1/(\beta d)$, the rescaled curves at different $V$ lie remarkably well on top of each other; in particular, since there is a large factor of more than $10^3$ between the smallest and the largest volume. 

To quantify the values of the exponents and their errors we make use of the fit routine employed in Ref.~\cite{Orioli:2015dxa} and refer to appendix~\ref{sec:FitProcedure} for more details. This yields
\begin{align}
	\frac{\alpha}{\beta d} =&\, 1.00 \pm 0.02 \, , 
	\label{eq:alphabetadelta} \\
	\frac{1}{\beta d} =&\, 0.57 \pm 0.03 \, ,
	\label{eq:betadelta} 
\end{align}
where the error bars are due to statistical averaging and fitting errors. We emphasize again that all our numerical values are obtained from simulations in $d=3$ spatial dimensions. Nevertheless, we keep here the parameter $d$ in the notation to reflect the fact that from the scaling ansatz (\ref{eq:Fscaling}) only the combination $\beta d$ of the scaling exponent for central time ($\beta$) and for volume ($d$) enters.

We are now going to extract the value of $z/d$ from (\ref{eq:FscalingV}) for $\Delta t\neq0$. For visualization purposes, we plot in Fig.~\ref{fig:selfsim_fit2} the rescaled correlation function $F_V$ as a function of $V^{-z/d}\Delta t$ for different values of $V^{-1/(\beta d)}\tau$. To establish the scaling behavior requires the comparison of the correlation function for different volumes $V_i$ at different times $\tau_i$, when plotted versus $\Delta t$. In particular, the times chosen need to fulfil $\tau_i/\tau_j=(V_i/V_j)^{1/(\beta d)}$. In doing so, one needs to make sure that the times $\tau_i$ lie within the regime where scaling is valid, which lasts longer for larger volumes according to (\ref{eq:scalingtaustar}) and (\ref{eq:scalingdeltatstar}). For instance, we find that for $256^3$ lattice sites the scaling regime is approximately given by the range of times $t,t'\in[200,3000]$ and for $512^3$ it is $t,t'\in[200,7000]$. Therefore, we plot in all figures values of $\tau$ and $\Delta t$ which lie approximately within these intervals.

One observes from Fig.~\ref{fig:selfsim_fit2} that the rescaled curves lie pairwise on top of each other to remarkable accuracy. This is the first demonstration of scaling dynamics in unequal-time correlation functions close to the nonthermal fixed point. Although we show only a couple of different times, we note that the agreement is valid for the whole scaling regime. In order to extract the exponents, we use our previous result (\ref{eq:betadelta}) and employ the fit routine to obtain
\begin{align}
	\frac{z}{d} =&\, 0.61 \pm 0.05 \, .
	\label{eq:zdelta} 
\end{align}
As a consistency check, we find that the result for $z/d$ does not depend much on whether we fix both $\alpha/(\beta d)$ and $1/(\beta d)$ by (\ref{eq:alphabetadelta}) and (\ref{eq:betadelta}) or only one of them when applying the fit routine to extract exponents.

\begin{figure*}[t!]
	\centering
	\includegraphics[width=1.0\textwidth]{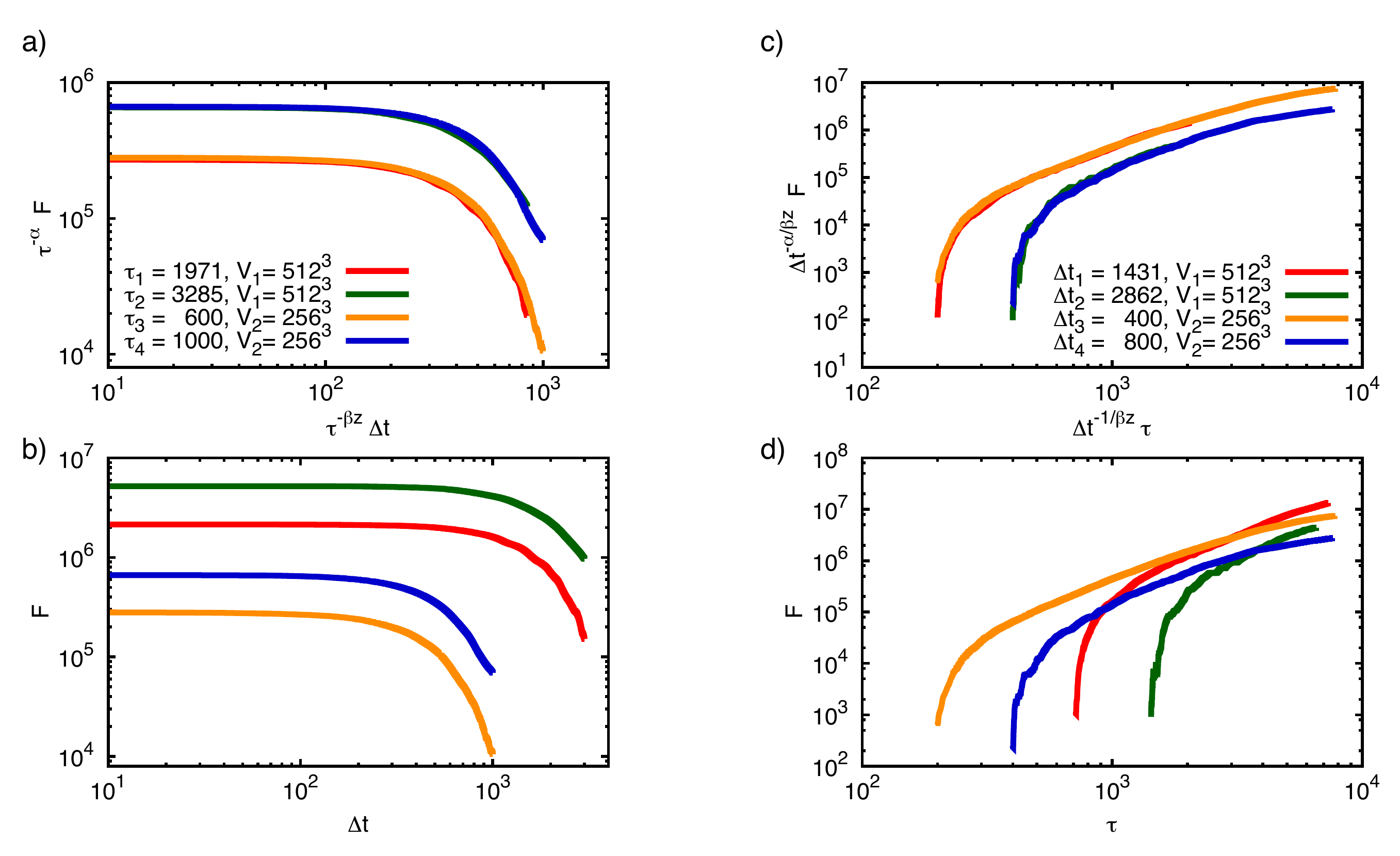}
	\caption{a) $F_\tau=\tau^{-\alpha}\, F$ as a function of the rescaled relative time $\tau^{-\beta z}\Delta t$ for two sets of values of $\tau^{-\beta d} V$ employing the exponents $\alpha=1.74$ and $\beta z=1.07$. b) Unrescaled function $F(\tau,\Delta t,V)$ for given values of $\tau$ as a function of $\Delta t$. c) $F_{\Delta t}=\Delta t^{-\alpha/\beta z}\, F$ as a function of the rescaled central-time $\Delta t^{-1/\beta z}\tau$ for given sets of $\Delta t^{-d/z} V$ with the same exponents. d) Unrescaled function $F(\tau,\Delta t,V)$ as a function of $\tau$ for given values of $\Delta t$.}
	\label{fig:selfsim_fit3}
\end{figure*}

We can do the same type of analysis using the scaling forms (\ref{eq:Fscalingtau}) or (\ref{eq:Fscalingdt}), which leads to a determination of the same exponents, however, in different combinations.
Fig.~\ref{fig:selfsim_fit3}a shows the correlation function $F_\tau =\tau^{-\alpha}\, F$ as a function of $\tau^{-\beta\, z} \Delta t$
for two sets of values of $\tau^{-\beta d} V$ with the volumes $256^3$ and $512^3$. The original function $F$ without rescalings is given in Fig.~\ref{fig:selfsim_fit3}b for comparison. The rescaled curves lie again well on top of each other. With the value of $\beta d$ given by (\ref{eq:betadelta}) one obtains from the fit routine
\begin{align}
\alpha =&\, 1.74 \pm 0.03 \, ,
\label{eq:alpha} \\
\beta z =&\, 1.07 \pm 0.06 \, ,
\label{eq:betaz}
\end{align}
which are consistent with the previous results within errors.

Fig.~\ref{fig:selfsim_fit3}c shows $F_{\Delta t} =\Delta t^{-\alpha/\beta z}\, F$ as a function of the rescaled central-time $\Delta t^{-1/\beta z}\tau$ for given sets of $\Delta t^{-d/z} V$ with the exponents found above. For comparison, Fig.~\ref{fig:selfsim_fit3}d displays the correlation function $F (\tau, \Delta t,V )$ for given values of $\Delta t$ versus the central-time coordinate $\tau$ without rescalings. The curves corresponding to different volumes $V_i$ and fulfilling $(\Delta t_i/\Delta t_j)^{d/z}=V_i/V_j$ lie pairwise well on top of each other. We checked that the results one obtains for exponents are consistent with the ones presented above within errors. For the plots one needs to fix the value for the dynamical scaling exponent $z$ in order to determine the values of fixed $\Delta t$ in different volumes. Furthermore, in Figs.~\ref{fig:selfsim_fit2} and \ref{fig:selfsim_fit3} we only use the largest volumes with $256^3$ and $512^3$ lattice points since the smaller available volumes are not in the scaling regime for relevant times. Nevertheless, we checked that comparing with data for $128^3$ and $256^3$ lattices one gets similar results, although they are less reliable due to the short duration of the scaling regime.

The above values for the universal scaling exponents
along with the scaling functions displayed represent our central results. In order to interpret them, we first note that  
the scaling relation $\alpha = \beta d$ reflects particle transport~\cite{Orioli:2015dxa}, which according to (\ref{eq:alphabetadelta}) is well realized by the scaling solution observed. Since with (\ref{eq:alpha}) we have $\alpha > 0$ the particle transport occurs from short to long distance scales, which characterizes an inverse cascade in agreement with the analysis of equal-time correlation functions in Ref.~\cite{Orioli:2015dxa}. The inverse particle cascade leads to the formation of a Bose condensate~\cite{Orioli:2015dxa}.  

The result (\ref{eq:zdelta}) represents the first direct determination of the dynamic scaling exponent $z$ for this nonthermal fixed point. Setting $d=3$ we obtain $z=1.84\pm0.15$. This value clearly excludes a ``linear dispersion" ($z \rightarrow 1$) in this scaling regime, but appears marginally consistent with a quadratic one ($z\rightarrow 2$). As a consistency check, we note that practically the same value for $z$ is also obtained from (\ref{eq:betaz}) using (\ref{eq:betadelta}) for $d=3$ giving $\beta = 0.58 \pm 0.03$. In addition, (\ref{eq:betaz}) conveys the important information that $z$ is rather accurately determined by $1/\beta$, even though the result for the errors stated indicates a small deviation. The agreement of $z$ and $1/\beta$ is, e.g., assumed in related studies of equal-time correlators in Refs.~\cite{SachdevPhysRevA.54.5037,karl2016strongly}.

Since the errors reflect only statistical uncertainties and the accuracy of the fit procedure, systematic errors could increase the error bars somewhat. To get an idea about possible systematic errors, we note that in Ref.~\cite{Orioli:2015dxa} the values for $\alpha$ and $\beta$ were obtained from the scaling behavior of a momentum distribution function. In this work we extract exponents from the (un)equal-time scaling of a volume-averaged quantity (\ref{eq:n_def_stat_nonrel}), which reflects properties of the correlator at zero spatial momentum only. 

If we repeat, for comparison, the momentum scaling analysis of Ref.~\cite{Orioli:2015dxa} for the distribution function with our current numerical setup, we obtain $\alpha \rightarrow 1.64\pm 0.16$ and $\beta \rightarrow 0.55 \pm 0.02$ consistent with Ref.~\cite{Orioli:2015dxa}. The relatively large error for $\alpha$ with this fit procedure is a result of the rather weak dependence of the distribution function at low momenta, and thus less accurate than our result (\ref{eq:alpha}), which is explained in more detail in the appendix~\ref{sec:FitProcedure}. In comparison, the value for $\beta$ obtained in this way has relatively small statistical errors and comes out directly from the fit procedue, i.e.,~without involving products as $\beta z$ or $\beta d$. Plugging this value naively into (\ref{eq:betaz}), or even into (\ref{eq:betadelta}) with (\ref{eq:zdelta}) treating the $d$ from the scaling ansatz (\ref{eq:Fscaling}) as an independent parameter,\footnote{
Such a procedure would lead, for instance, from (\ref{eq:betadelta}) and (\ref{eq:beta}) to the value
	\begin{equation}
	d \rightarrow 3.18 \pm 0.13 \, ,
	\label{eq:delta}
	\end{equation}
	for the scaling parameter $d$ in (\ref{eq:Fscaling}). The deviation from the spatial dimension three, for the
	statistical and fit error given, may point to a moderate additional systematic error. 
}
would lead to $z \rightarrow 1.94 \pm 0.11$. While this is still fully consistent with our above result for $z$, its somewhat higher central value might be viewed as an indication for a possible quadratic dispersion.    

The discussion about the deviation from a quadratic dispersion relation is also closely related to the question of a non-vanishing anomalous dimension $\eta$ describing the deviation from canonical scaling~\cite{Orioli:2015dxa}, as recently addressed also in two spacial dimensions using equal-time correlations~\cite{karl2016strongly}. Following Ref.~\cite{Orioli:2015dxa} employing large-$N$ expansions, the anomalous dimension may be determined by the relation
\begin{equation}
\beta = \frac{1}{2-\eta} \, .
\label{eq:betaeta} 
\end{equation}  
Taking the (somewhat more accurate) value of $\beta$ obtained from a fit to momentum scaling distributions as explained above, we get
\begin{equation}
\eta = 0.19 \pm 0.08 \, .
\label{eq:eta}
\end{equation}
The smallness of the anomalous dimension makes it difficult to draw definite conclusions in view of the relatively large error bars. However, the central value obtained for $\eta$ at the nonthermal fixed point is rather large if compared to typical values of the corresponding thermal critical exponent, which is on the order of a few percent in scalar theories in three dimensions.

\section{Conclusion}\label{sec:conclusions}

In this work we have presented first results on universal scaling exponents and scaling functions for unequal-time correlation functions describing nonthermal fixed points. In particular, this allows us to directly establish the value of the dynamic scaling exponent $z$, characterizing the frequency dependence of unequal-time correlations or the dispersion, and its close relation to $1/\beta$ describing the scaling of equal-time quantities such as the distribution function.  

The method we have employed is based on a systematic finite-size scaling analysis of classical-statistical simulations for an interacting complex scalar field theory in three spatial dimensions. Since the system has a finite size, we are able to quantify the scaling of the characteristic time scales $\tau_*$ for condensation and of the correlation time $\Delta t_*$ with volume. Since the former scales $\sim V^{1/(\beta d)}$ and the latter $\sim V^{z/d}$, the established similarity between the exponents $1/\beta$ and $z$ entails a corresponding scaling of both condensation and correlation times.  

To put these results into context, we note that also the corresponding relativistic model belongs to the same universality class~\cite{Orioli:2015dxa}. In all these theories, the infrared scaling behavior is part of a dual cascade, with a turbulent energy cascade towards shorter distances~\cite{Micha:2004bv}. In particular, for the scaling properties of the direct energy cascade there is no such similarity between the corresponding values of $z$ and $1/\beta$, which even turn out to have opposite signs in that case~\cite{Orioli:2015dxa}. In this respect, the non-perturbative inverse particle cascade and the perturbative direct energy cascade are found to behave very differently.

Since the observed value of $z$ close to two makes it rather difficult to distinguish it from several other known universality classes, we emphasize that the universal shape of the scaling forms we have computed provides important additional information. For instance, it has been analyzed in great detail already in Refs.~\cite{Berges:2010ez,Orioli:2015dxa,Moore:2015adu} that the shape of the momentum scaling functions obtained from equal-time correlation functions exhibits a remarkable universality across $N$-component scalar field theories with different $N$. Since the complex scalar theory we are considering corresponds to $N=2$ real scalar field components, we expect for the unequal-time scaling functions a similar universality for different values of $N$ to hold as for the equal-time functions. This is supported also by the close relation between equal- and unequal-time scaling exponents that we established in this work. This can be used to distinguish the scaling behaviour, e.g., from coarsening phenomena. The latter strongly reflect the topological obstructions that depend on $N$. 

The remarkably large universality class associated to the nonthermal fixed point is rooted in the extreme far-from-equilibrium situation of very high typical excitations or occupation numbers. Since the characteristic occupancies are non-perturbatively large, $\sim 1/\zeta \gg 1$ in the dilute regime, they can become insensitive to the details of the underlying thermal or vacuum structure for which typical occupancies are of order unity.

While these extreme conditions may appear unnatural at first sight, we emphasize that these are attractor solutions: there is no relevant parameter to tune, as for instance the tuning of a critical temperature to be at a thermal transition. Moreover, the extreme conditions appear in situations associated to nonequilibrium instabilities in a wide range of applications from particle-physics cosmology to condensed matter physics. The universality opens, therefore, the exciting possibility to learn something about the early stages of our universe from table-top experiments with, e.g., ultracold atoms.

\begin{acknowledgments}
We thank Kirill Boguslavski for very helpful discussions and collaborations on related work. This work was supported in part by the ExtreMe Matter Institute EMMI at the GSI Helmholtzzentrum f\"ur Schwerionenforschung, Darmstadt. This work is part of and supported by the DFG Collaborative Research Centre ``SFB 1225 (ISOQUANT)". 
\end{acknowledgments}

\appendix

\section{Numerical fit procedure}\label{sec:FitProcedure}

In this section we describe the fit routine used to quantify the central values and statistical errors of the scaling exponents given in the main text. We use for this the self-similar scaling behavior of the two-point unequal-time correlation function according to the scaling forms (\ref{eq:FscalingV})--(\ref{eq:Fscalingdt}), which depend on different combinations of the exponents $(\alpha,\beta, z)$ and on $d$. Although $d$ is associated to the fixed dimension of the system, here we keep the discussion more general by treating it as an independent parameter. At the end of the section, we also give some details about the scaling of the equal-time correlator, which were not given in the main text.

The fit procedure is based on the study of equal-time correlators in Ref.~\cite{Orioli:2015dxa} and we extend it here to unequal-time correlation functions. Due to the number of exponents and different ways to write the scaling forms, we divide our analysis into four steps where different combinations of exponents are computed. We start by considering the scaling ansatz (\ref{eq:FscalingV}), which will serve to exemplify the general strategy of our fit routine. To numerically quantify the deviation from the self-similar evolution, we need to compare the correlation function $F(\tau,\Delta t,V)$ at different volumes by appropriately rescaling $\tau$ and $\Delta t$. Having this in mind, we define the rescaled correlation function
\begin{align}
 F_{\textit{resc}}(\tau,\Delta t,V)&\equiv\left (V/V_{\textit{ref}} \right )^{-\alpha/(\beta d)} \label{eq:rescF_def}\\ 
 & \times F\left((V/V_{\textit{ref}})^{1/(\beta d)}\,\tau, (V/V_{\textit{ref}})^{z/d}\, \Delta t, V\right)\nonumber \, ,
\end{align}
where $V_{\textit{ref}}$ is some reference volume to which we compare.
Using this definition, the self-similar scaling (\ref{eq:FscalingV}) can be rewritten as $F_{\textit{resc}}(\tau,\Delta t,V)= F(\tau, \Delta t, V_{\textit{ref}})$. Hence, deviations from scaling at a given point are given by
\begin{equation}
\Delta F(\tau,\Delta t,V)\equiv F_{resc}(\tau,\Delta t,V)-F(\tau,\Delta t,V_\textit{ref}) \, .
\label{eq:deltaF_def}
\end{equation}
Using (\ref{eq:deltaF_def}) we will define a $\chi^2$-function which adds up, with the appropriate weight, all the deviations $\Delta F$ over a given range of $\tau$ or $\Delta t$. This $\chi^2$-function quantifies, thus, the total deviation from self-similarity which we will try to minimize by a suitable choice of exponents.

\begin{figure}[t!]
\centering
\includegraphics[width=.49\textwidth]{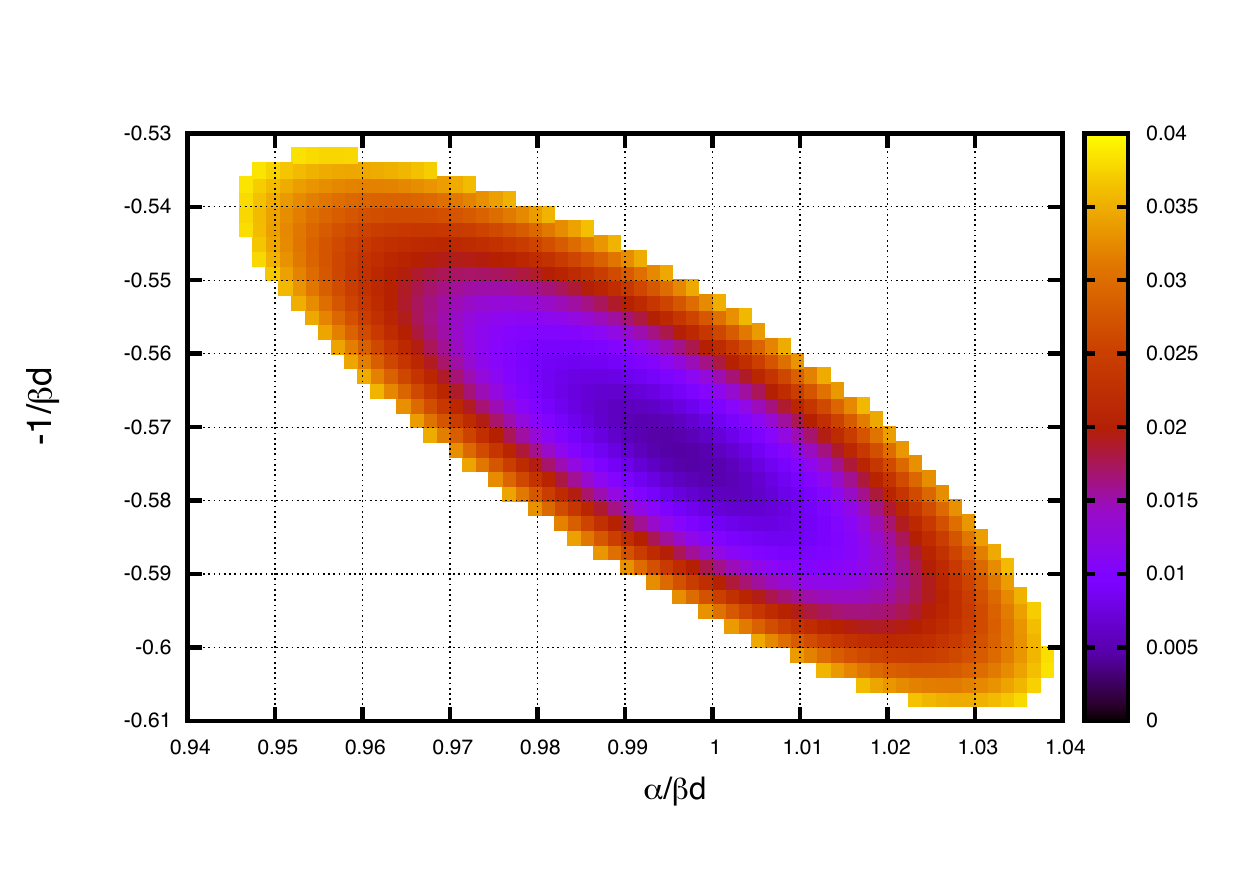}
\caption{The parameter $\chi^{2}(\alpha/(\beta d),1/(\beta d),V_{\textit{ref}})$ as a function of the exponents $\alpha/(\beta d)$ and $1/(\beta d)$ with the reference volume $V_{\textit{ref}}=256^{3}$ and the fit interval $\tau \in [50,8000]$.}\label{fig:ChiFS1}
\end{figure}

To be more specific, we consider our first fit scheme at equal times, i.e.~$\Delta t=0$ ($\tau=t$). This allows us to consider just the pair of exponents $\alpha/(\beta d)$ and $1/(\beta d)$. Using (\ref{eq:deltaF_def}) we define for given $V$ and $V_{\textit{ref}}$ :
\begin{equation}
\chi^{2} \biggl (\dfrac{\alpha}{\beta d},\dfrac{1}{\beta d}\biggl ) \equiv \int \biggl ( \dfrac{\Delta F(\tau,0,V)}{F(\tau,0,V_{\textit{ref}})}\biggl )^{2}\frac{\mathrm{d} (\log(\tau))}{\mathbb{T}} \, ,
\label{eq:Vol_ChiSquare} 
\end{equation}
where the integration limits are chosen to be within the self-similar regime and
\begin{equation}
\mathbb{T}\equiv \int \dif (\log(\tau))
\end{equation}
is the normalization of the integral.
Due to the power-law nature of the correlation function in the scaling regime, we integrate over $\log(\tau)$ with $\tau>0$. This enhances the sensitivity of the integral to small times $\tau$, where the density of points is smaller.
For this first fit scheme we have considered the set of volumes $\lbrace 32^{3},64^{3},128^{3},256^{3},512^{3}\rbrace$ and have chosen $V_{\textit{ref}}$ to be $128^{3}$. Each volume $V$ of this set is compared to $V_{\textit{ref}}$ individually and then the $\chi^2$ of the different volumes are added up.
Varying the values of the exponents $\alpha/(\beta d)$ and $1/(\beta d)$, we obtain the distribution of $\chi^2$ values shown in Fig.~\ref{fig:ChiFS1}. The set of exponents $\{ (\alpha/(\beta d))^{*} , (1/(\beta d))^{*} \}$ that minimizes the $\chi^2$-function (see dark shaded area) is the one that makes the rescaled curves lie most accurately on top of each other (see Fig.~\ref{fig:selfsim_fit1}) and hence constitutes our final result.

From the width of the distribution we extract the statistical errors. For this we first define a likelihood function 
\begin{equation}
W\biggl [\dfrac{\alpha}{\beta d},\dfrac{1}{\beta d}\biggl ]=W_{0}^{-1} \exp\biggl (-\frac{\chi^{2}(\alpha/(\beta d),1/(\beta d))}{\chi^{2}_{min}}\biggl )
\end{equation}
where $\chi^2_{\textit{min}} \equiv \chi^2((\alpha/(\beta d))^{*} , (1/(\beta d))^{*})$ is the minimal value of $\chi^2$ and the normalization constant $W_{0}$ is chosen such that
\begin{equation}
\int\,W \biggl [\dfrac{\alpha}{\beta d},\dfrac{1}{\beta d}\biggl ] \dif \biggl (\frac{\alpha}{\beta d}\biggl ) \dif \biggl ( \frac{1}{\beta d}\biggl )=1.
\end{equation} 
By integrating over only one of the exponents, we obtain marginal likelihood functions $W[\alpha/(\beta d)]$ and $W[1/(\beta d)]$ defined via
\begin{equation}
W\biggl [ \frac{\alpha}{\beta d} \biggl ]\equiv\int\, W \biggl [\dfrac{\alpha}{\beta d},\dfrac{1}{\beta d}\biggl ] \dif \biggl ( \frac{1}{\beta d}\biggl ) \, ,
\end{equation}
and
\begin{equation}
W \biggl [ \frac{1}{\beta d} \biggl ]\equiv\int\,W \biggl [\dfrac{\alpha}{\beta d},\dfrac{1}{\beta d}\biggl ] \dif \biggl (\frac{\alpha}{\beta d}\biggl ) \, .
\end{equation}
Approximating these two functions with Gaussian distributions, we obtain the errors of $(\alpha/(\beta d))^{*}$ and $(1/(\beta d))^{*}$ from the standard deviation \cite{Orioli:2015dxa}. In this way, we get the central value and statistical errors of the exponents given in (\ref{eq:alphabetadelta}) and (\ref{eq:betadelta}).

In our second fit scheme we consider the same scaling form (\ref{eq:FscalingV}) but for unequal times $\Delta t\neq0$. To apply the above fit routine, we consider slices of constant volume $V$ and central-time $\tau$ as a function of the remaining variable $\Delta t$. We consider in this case the set of volumes $\lbrace 128^{3}, 256^3, 512^{3}\rbrace$ and set $V_{\textit{ref}}$ to be $256^3$. Due to the rescaling of (\ref{eq:rescF_def}), the central times $\tau$ of the different volumes $V$ have to fulfil
\begin{equation}\label{eq:SecondFit_RescaledTime} 
\dfrac{\tau}{\tau_{\textit{ref}}}=\left (\dfrac{V}{V_{\textit{ref}}}\right )^{1/(\beta d)} \, ,
\end{equation}
where $\tau_{\textit{ref}}$ corresponds to $V_{\textit{ref}}$. It is important to note that both $\tau$ and $\tau_{\textit{ref}}$ have to lie within the scaling regime. For $V_{\textit{ref}}$ we choose $\tau_{\textit{ref}}\in[600,2000]$. To solve for $\tau$ in (\ref{eq:SecondFit_RescaledTime}) we use the exponent $1/(\beta d)$ that was determined by the first fit (\ref{eq:betadelta}). Hence, only $\alpha/(\beta d)$ and $z/d$ remain as fitting parameters. Using this we define again a $\chi^2$ function to be minimized for fixed $V$, $V_{\textit{ref}}$, $\tau$ and $\tau_{\textit{ref}}$ by
\begin{equation}\label{eq:Vol_ChiSquare_DT} 
\chi^{2}\biggl (\dfrac{\alpha}{\beta d},\dfrac{z}{d}\biggl ) \equiv \int \biggl ( \dfrac{\Delta F(\tau,\Delta t,V)}{F(\tau_{\textit{ref}},\Delta t,V_{\textit{ref}})}\biggl )^{2}\dfrac{\dif (\log(\Delta t))}{\mathbb{D}}
\end{equation}
with the normalization factor 
\begin{equation}
\mathbb{D}\equiv \int \dif (\log(\Delta t)) \, .
\end{equation}
The central values and statistical errors of the exponents $\alpha/(\beta d)$ and $z/d$ are obtained from $\chi^2$ in the same way as presented above. To obtain more accurate results, we have considered around $30$ different $\tau_{\textit{ref}}$ within the scaling regime and averaged the final result over all fits. As explained in the main text, we can only compare two different volumes with each other because of the condition (\ref{eq:SecondFit_RescaledTime}) and the requirement that all times lie inside the scaling regime. Therefore, the results given in (\ref{eq:zdelta}) and used in Fig.~\ref{fig:selfsim_fit2} correspond to the largest volumes $256^3$ and $512^3$.

An important point concerns the error of the exponent $1/(\beta d)$ which propagates into the chosen values of $\tau$ by means of (\ref{eq:SecondFit_RescaledTime}). To quantify this additional source of error, we first use (\ref{eq:SecondFit_RescaledTime}) to define the times $\tau^\pm$ lying at the edges of the error interval by
\begin{equation}\label{eq:ErrorTimes} 
\dfrac{\tau^{\pm}}{\tau_{\textit{ref}}}=\left (\dfrac{V}{V_{\textit{ref}}}\right )^{\frac{1}{\beta d}\pm\Delta\bigl (\frac{1}{\beta d}\bigl ) } \, ,
\end{equation}
where $\Delta\bigl (1/\beta d\bigl )$ denotes the error of the exponent $1/(\beta d)$. Repeating the fit routine with the values $\tau^{\pm}$ yields slightly different central values for the exponents $\alpha/(\beta d)$ and $z/d$. We interpret the deviation from our main result, calculated with $\tau$ from (\ref{eq:SecondFit_RescaledTime}), as the error propagated from $\Delta(1/\beta d)$. Adding this extra error quadratically to the statistical error obtained before from the width of the $\chi^2$-distribution, we obtain the final result given by (\ref{eq:zdelta}) and $\alpha/(\beta d)=1.0\pm 0.07$, which is consistent with (\ref{eq:alphabetadelta}).

\begin{figure}[t!]
	\centering
	\includegraphics[width=.5\textwidth]{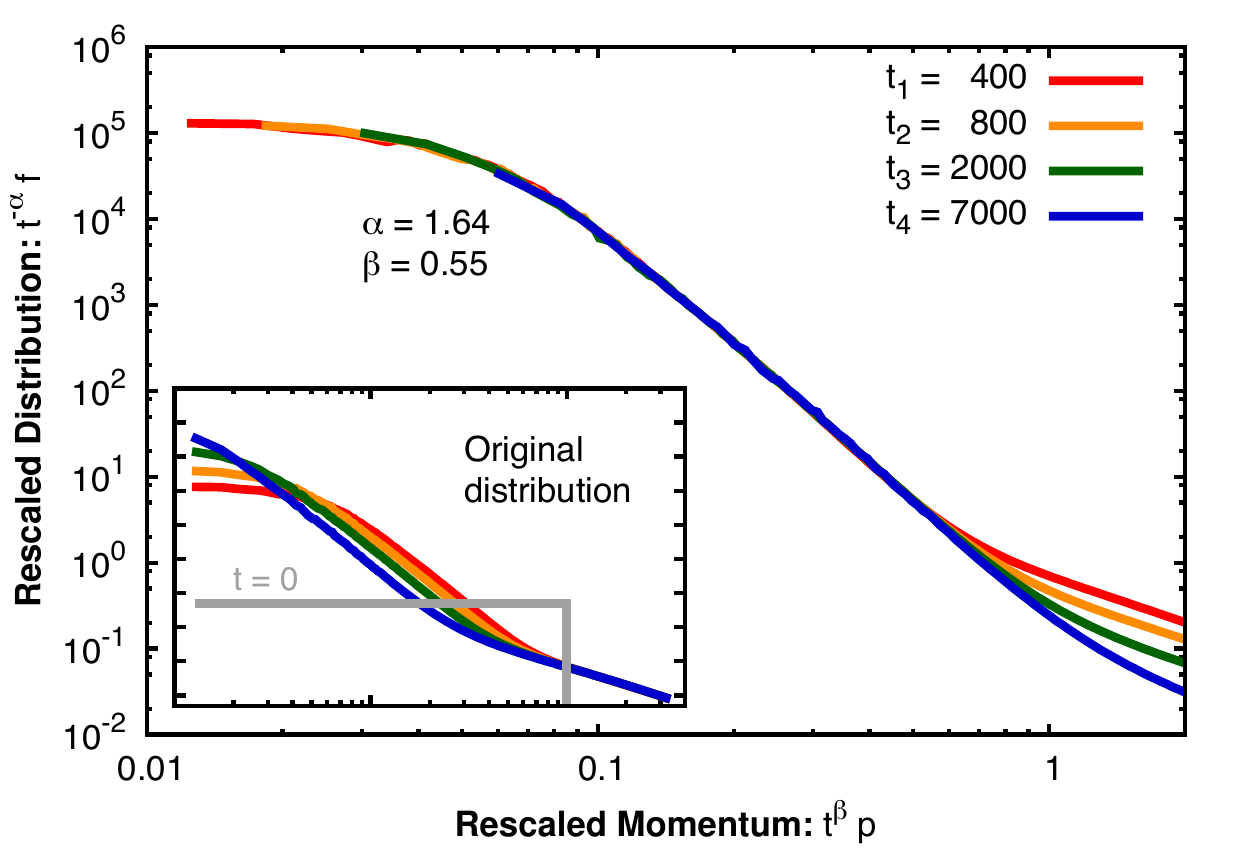}
	\caption{Rescaled distribution function $\tau^\alpha f(\tau,\mbf p)$ defined in Ref.~\cite{Orioli:2015dxa} as a function of the rescaled momentum $\tau^\beta \mbf p$ with the exponents $\alpha=1.64$ and $\beta=0.55$ for $512^3$ lattice points. The inset shows the original function without rescaling.}
	\label{fig:selfsim_alphabeta}
\end{figure}

Our third step consists in considering the scaling form (\ref{eq:Fscalingtau}). The fit routine follows along the same lines as before except for a different rescaled correlation function given by
\begin{align}
F_{\textit{resc}}(\tau,\Delta t,V)&\equiv\left (\tau/{\tau_{\textit{ref}}}\right )^{-\alpha}\, F\bigl (\tau, (\tau/\tau_{\textit{ref}})^{\beta z}\, \Delta t,\nonumber\\
&\quad (\tau/\tau_{\textit{ref}})^{\beta d}\, V\bigl ) \, .
\end{align}
The $\chi^{2}$-function can be defined similarly to \eqref{eq:Vol_ChiSquare_DT}. The main difference is that we rescale with the central-time coordinate instead of the volume. Volumes and central times have to be chosen again such that \eqref{eq:SecondFit_RescaledTime} is fulfilled. We choose $V_{\textit{ref}}=256^{3}$ with $\tau_{\textit{ref}}\in[600,2000]$ and compare to $V=512^{3}$ with $\tau$ from (\ref{eq:SecondFit_RescaledTime}), as a function of $\Delta t$. Proceeding as before yields the fit results (\ref{eq:alpha}) and (\ref{eq:betaz}) where the error in $\tau$ coming from \eqref{eq:ErrorTimes} was taken into account as above. In an analogous way, one can check the consistency of our results in a fourth step by making use of the scaling relation \eqref{eq:Fscalingdt}.

\begin{figure}[t!]
\centering
\includegraphics[width=.49\textwidth]{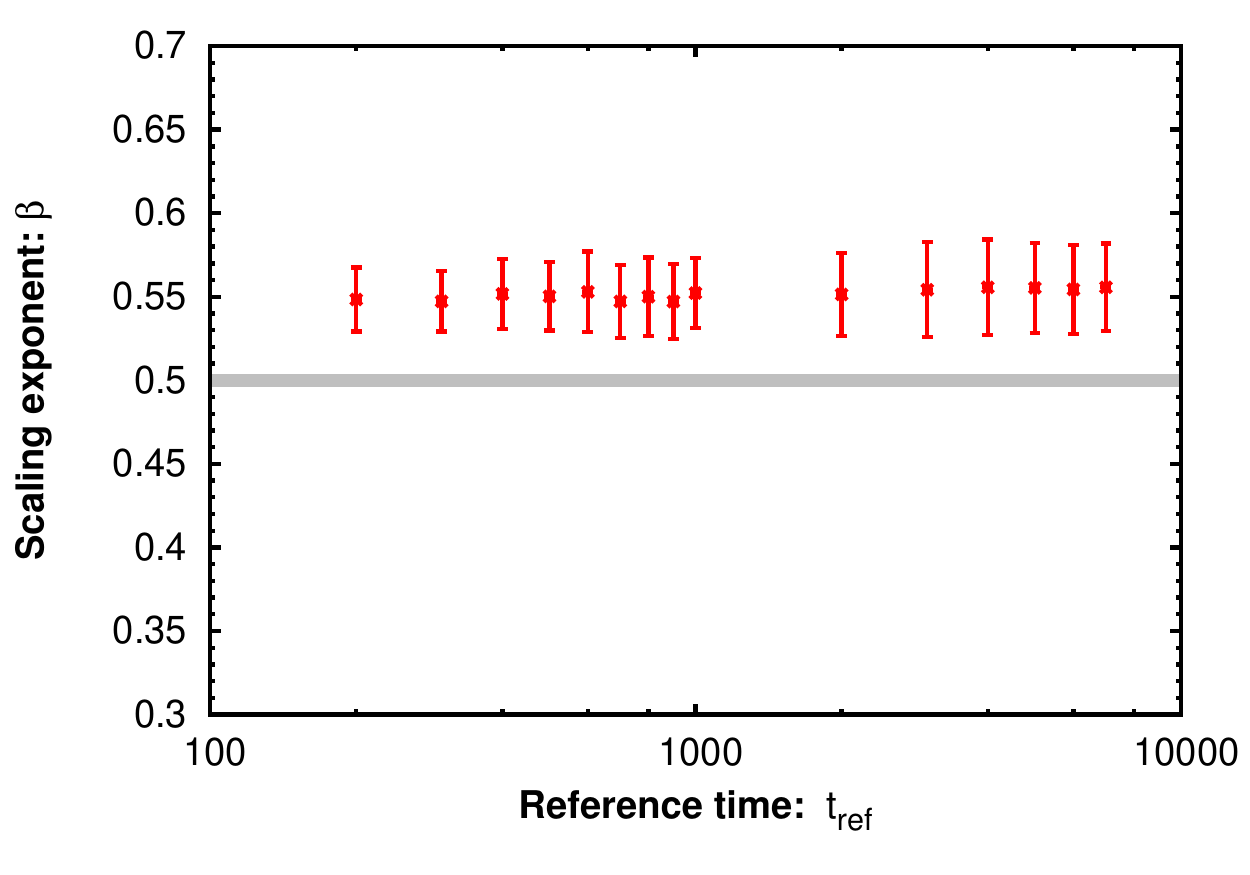}
\caption{Fit result for the exponent $\beta$ with different values of the reference time $t_{\textit{ref}}$ from the equal-time distribution function. It demonstrates that the value is very stable within the considered range of $t_{\textit{ref}}$.}
\label{fig:Beta}
\end{figure}

While the application of the fit procedure to unequal-time correlation functions at zero momentum has been discussed in the main text, here we give some more details about the application to momentum-dependent equal-time correlators. The fit routine can be adapted straightforwardly \cite{Orioli:2015dxa} to study the self-similar behavior of the distribution function $f(t,\mathbf{p})$, defined for homogeneous and isotropic systems as the spatial Fourier transform of $F(t,t,\mbf x-\mbf x')$ given by \cite{Orioli:2015dxa}. The distribution function $f(t,\mbf p)$ evolves in the universal regime as
\begin{equation}
	f(t,\mbf p) = t^\alpha f_S(t^\beta \mbf p) \, ,
\label{eq:selfsim_equaltime}
\end{equation}
where $f_S(\mbf p)=f(1,\mbf p)$. This scaling form gives us access to $\alpha$ and $\beta$ separately. In general, the scaling ansatz for $f(t,\mbf p)$ should also include the dependence on the volume. However, our numerics reveal that this quantity is rather insensitive to changes of the volume during the universal regime (see also the inset of Fig.~\ref{fig:selfsim_fit1}). To extract the central values and errors of the exponents $\alpha$ and $\beta$, one compares curves at different times $t$ with a reference time $t_{\textit{ref}}$ and computes a $\chi^2$-function in analogy to the procedure outlined above. In this way, we get
\begin{align}
	\beta = 0.55 \pm 0.02 \, ,
\label{eq:beta}
\end{align}
and $\alpha=1.64\pm 0.16$, which is consistent with the results of Ref.~\cite{Orioli:2015dxa}. Fig.~\ref{fig:selfsim_alphabeta} shows how curves corresponding to different times lie on top of each other after rescaling with these exponents, reflecting the self-similar evolution of $f(t,\mbf p)$. We note that the value of $\alpha$ obtained in this way is consistent within error bars with (\ref{eq:alpha}). The relatively large error for $\alpha$ obtained from the momentum-dependent analysis is related to the form of the distribution function, whose plateau at small momenta vanishes for long times (see Fig.~\ref{fig:selfsim_alphabeta}).

\begin{figure}[t!]
	\centering
	\includegraphics[width=.5\textwidth]{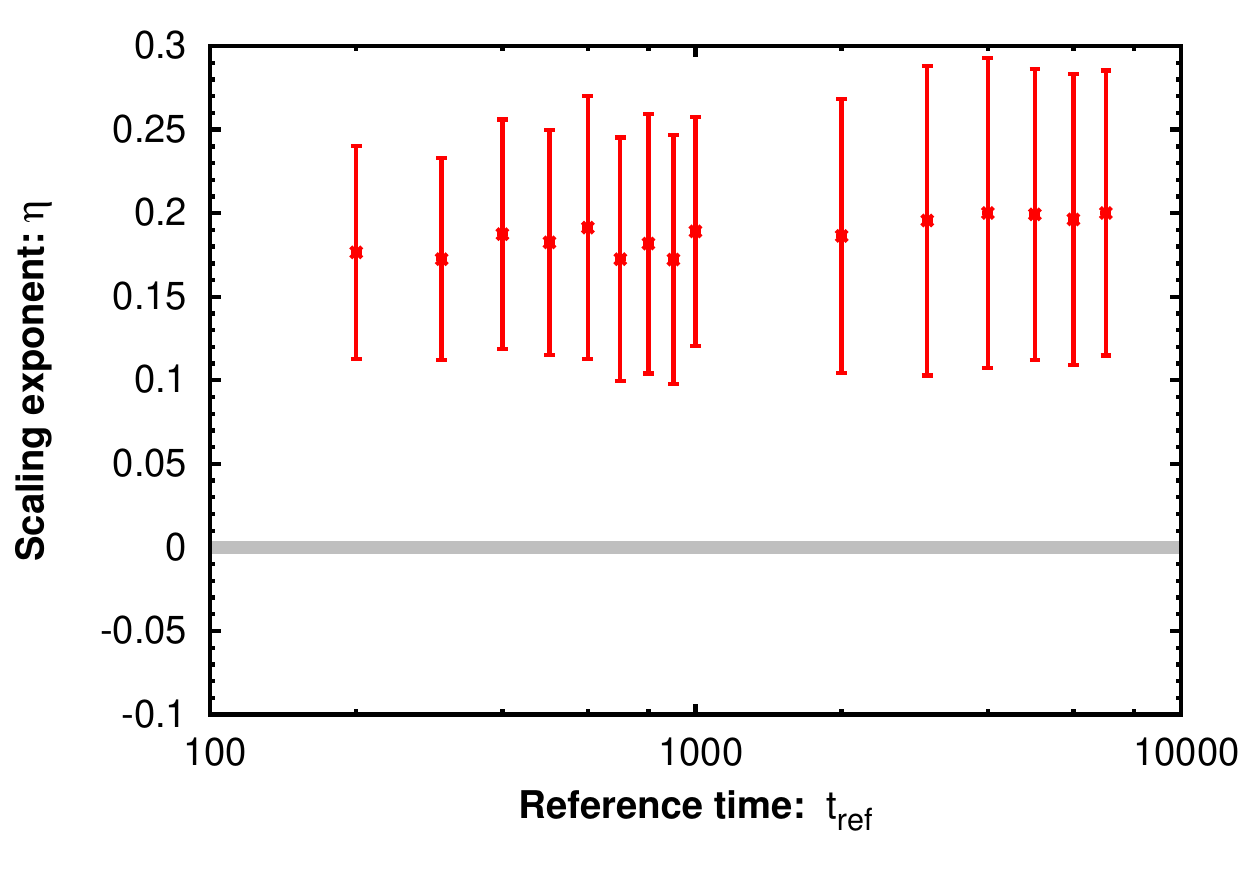}
	\caption{Anomalous dimension $\eta$ as obtained from the relation $\beta=1/(2-\eta)$ using large-$N$ techniques~\cite{Orioli:2015dxa}. The exponent $\beta$ is obtained from the scaling of $f(t,\mbf p)$. The time interval chosen for the scaling analysis starts with $t_{\text{ref}}$.}
	\label{fig:eta}
\end{figure}

As discussed in the main text, we can use the large-N result (\ref{eq:betaeta}) to relate the anomalous dimension $\eta$ to the value of $\beta$. For this we vary the value of the reference time $t_{\text{ref}}$ for $512^3$ within $[200,7000]$ and plot the obtained exponent in Fig.~\ref{fig:Beta}. As one can see, the value of $\beta$ obtained is rather stable over the whole universal regime. Using (\ref{eq:betaeta}), we plot the corresponding values of the anomalous dimension $\eta$ in Fig.~\ref{fig:eta}. While the error bars are rather large, the central value is approximately constant over the whole range and indicates a deviation from zero.

\clearpage

\bibliography{mybibliography_UnequalTime}

\end{document}